\documentstyle[12pt]{article}
\def\a{\alpha}

\def\b{\beta}

\def\c{\gamma}
\def\d{\delta}
\def\e{\epsilon}

\def\l{\lambda}

\def\om{\omega}
\def\Om{\Omega}

\def\s{\sigma}
\def\t{\tau}
\def\th{\theta}

\def\beq{\begin{equation}}\def\eeq{\end{equation}}
\def\beqa{\begin{eqnarray}}\def\eeqa{\end{eqnarray}}
\def\barr{\begin{array}}\def\earr{\end{array}}

\let\bm=\bibitem
\def\nn{\nonumber}
\def\bd{\begin{document}}
\def\ed{\end{document}}
\def\ba{\begin{array}}
\def\ea{\end{array}}
\def\bea{\begin{eqnarray}}
\def\eea{\end{eqnarray}}
\def\ft#1#2{{\textstyle{{\scriptstyle #1}\over {\scriptstyle #2}}}}
\def\fft#1#2{{#1 \over #2}}
\newcommand{\be}{\begin{equation}}
\newcommand{\ee}{\end{equation}}
\newcommand{\eq}[1]{(\ref{#1})}
\def\eqs#1#2{(\ref{#1}-\ref{#2})}
\def\det{{\rm det\,}}
\def\tr{{\rm tr}}
\newcommand{\ho}[1]{$\, ^{#1}$}
\newcommand{\hoch}[1]{$\, ^{#1}$}
\def\ra{\rightarrow}
\def\uha{{\hat {\underline{\a}} }}
\def\uhc{{\hat {\underline{\c}} }}

\def\bR{{\bf R}}
\def\br{{\bf r}}
\def\bP{{\bf P}}
\def\bp{{\bf p}}
\def\bk{{\bf k}}
\def\bep{{\bf \epsilon}}
\def\bA{{\bf A}}

\def\De{\Delta}
\def\del{\partial}
\def\Gh{\hat{G}}

\def\ft{{\tilde{f}}}
\def\cM{ {\cal M} }

\newcommand{\auth}{\large C. K. Au\hoch{1,3}
and C. S. Chu\hoch{2,3} }

\begin{document}

\hfill{SISSA 1/98/FM}

\hfill{physics/9801020}


\vspace{20pt}

\begin{center}

{\Large\bf Finite Mass Effect on Two Photon Processes
in Hydrogenic Systems: Effective Scalar Photon Interaction}
\vspace{30pt}

\auth

\vspace{15pt}

\begin{itemize}
\item[$^1$] {\small \em Department of Physics and Astronomy,
University of South Carolina, Columbia, SC  29208, USA}
\item[$^2$] {\small \em
International School for Advanced Studies (SISSA),
Via Beirut 2, 34014 Trieste, Italy}
\item[$^3$] {\small \em Department of Physics, The Chinese University 
of Hong Kong, Hong Kong, China}
\end{itemize}

{\rm email: au@sc.edu, cschu@tao.fm.sissa.it}

\vspace{60pt}

{\bf Abstract}

\end{center}
 
We consider a hydrogenic system with a nucleus of finite mass. The
coupling of the radiation field to the center of mass motion gives rise
to an effective scalar type coupling. This induced scalar photon
interaction emerges as a correction in competition with the usual
multipole interactions. This effect is particularly important in
positronium where the electric quadrupole interaction is totally
suppressed. We illustrate this
effect with the two--photon decay  of metastable hydrogenic systems.

\pagebreak
\setcounter{page}{1}
\section{Introduction}

Recently, one of us \cite{au1} pointed out, within the framework of
nonrelativistic spinless QED, the emergence of an effective scalar photon
interaction for the atomic (hydrogenic) internal degree of freedom due
to finite mass corrections. In this simple model, the atom is a
composite system with two particles of finite masses ($m_1$ and $m_2$)
and charges ($Z_1$ and $Z_2$) of opposite sign. In contrast to the
relative motion which couples to the radiation field vectorially through
minimal coupling, the coupling of the center of mass motion to the
radiation field induces an effective scalar type coupling for the
relative motion. {\it Thus, when finite nuclear mass effect is taken
into account, the induced scalar photon interaction leads to a
modification of the multipole interactions of the atom with radiation.
}

Consider a system consisting of two particles of masses $m_1$, $m_2$ and
charges $Z_1 e, Z_2 e$.
Let $\bR$ and $\br$ denote the center of mass (c.m.) and the relative
coordinate vectors and $\bP$ and $\bp$ denote the momenta conjugate to
$\bR$ and $\br$. To be definite, we refer to particle 1 as the nucleus
and the particle 2 as the electron.
The finite mass correction is characterized by the
dimensionless quantity 
\be \d = m_2/M, \quad\quad \mbox{where $M =m_1 +m_2$} 
\ee 
is the total mass of the system.
In general, $0 \leq \d \leq 1$.
The infinite nuclear mass  limit corresponds to $\d=0$.
It has been shown \cite{au1} 
that the spinless non-relativistic QED Hamiltonian
describing this system is given by 
\be
H = H_0 + H_{int},
\ee
where
\bea
&&H_0= H_{2 particle} + H_{rad}, \nn\\
&&H_{int}= H_V +H_S + H_{sg}.
\eea
$H_0$ is the Hamiltonian describing the two-particle system ($H_{2
particle} $) and the radiation field ($H_{rad}$). 
$H_{int}$ describes the interaction between the two--particle system and
the radiation field. 
$H_V$ describes the coupling of the radiation to the relative 
motion and is given by the usual vector photon type interaction.
$H_{sg}$ is a seagull
type interaction responsible for simultaneous two--photon processes. 
$H_S$ describes the coupling of the radiation to the c.m. 
and is given by an effective scalar photon interaction Hamiltonian.
Explicitly, the forms of $H_V$, $H_S$, and  $H_{sg}$ are \cite{au1}:
\bea
&&H_V = \sum_{(\bep, \bk)}
\sqrt{\frac{2\pi}{\om}} e \bp \cdot \bep \, e^{i \bk \cdot \bR}
( \frac{Z_2}{m_2} e^{i (\d -1) \bk \cdot \br} - 
  \frac{Z_1}{m_1} e^{i \d \bk \cdot \br} ) \, a_{(\bep, \bk)} +h.c. \ ,
\label{HV} \\
&&H_S = -\sum_{(\bep, \bk)}
\sqrt{\frac{2\pi}{\om}} \frac{e \bP \cdot \bep}{M} \,
 e^{i \bk \cdot \bR} 
( Z_1 e^{i \d \bk \cdot \br} + 
  Z_2 e^{i (\d-1) \bk \cdot \br} ) \, a_{(\bep, \bk)} +h.c.  \ .
\label {HS} \\
&& H_{sg} =\frac{Z_1^2 e^2}{2 m_1} \bA(\br_1)^2 +
\frac{Z_2^2 e^2}{2 m_2} \bA(\br_2)^2  ,
\eea
where
\be
\bA(\br) =\sum_{(\bep, \bk)} \bep \sqrt{\frac{2\pi}{\om}} 
[e^{i \bk \cdot \br} a_{(\bep, \bk)} + h.c. \: ],
\ee
and $\bep$ is the polarization vector 
and $a_{(\bep, \bk)}$ is the annhiliation operator for the photon.

It has long been accepted that within nonrelativistic QED, a single
photon transition between two $S$-states ($L=0$) is impossible \cite{breit}. 
This
readily follows from parity consideration if only the vector interaction
is assumed. In \cite{au1}, the effects of finite mass correction on
one--photon transitions were studied. Taking into account the induced
scalar interaction, it was shown that the metastable $2S$ state can
decay by {\it one}--photon emission, which is otherwise forbidden if the
nucleus is taken to be infinitely massive. Being a finite mass
correction, this one photon decay rate is generally much smaller than
the two photon decay rate.  The M1 decay \cite{johnson}, 
though allowed relativisticaly,
is not considered here.

For a self conjugate system with $m_1=m_2$ and $Z_1 = -Z_2$,
it turns out that this finite mass correction effect vanishes
in the one--photon decay channel and so the leading
effect of finite mass correction will appear in two--photon processes.
It is the aim of this report to provide a systematic study of  
finite mass effects on two photon processes.

We now  give an estimate of the role of the effective scalar
photon interaction. For simplicity, 
we ignore the overall change in the state of the
center of mass motion, bearing in mind that a nonvanishing effective
scalar photon interaction requires a nonzero velocity for the center of
mass motion. But in the calculations  reported in sections 2 and 3 below, 
we do include the effects of the 
c.m. recoil.
As an example, we consider the two--photon decay of
positronium from the metastable $|2S>$ state to $|1S>$ state. 
In this case, $Z_1=-Z_2 =1$, 
$m_1=m_2=m_e$, $\d=1/2$ and $M=2 m_e$.
On expanding the exponential factors $exp( -i \bk \cdot \br /2)$ and 
$exp(-i \bk \cdot \br /2)$ in both eqns. \eq{HV} and \eq{HS}, we readily 
observe that all even electric multipole interactions are totally
suppressed in positronium \cite{post}. We now focus on the dominant
decay channels. For the E1 (electric dipole) channel arising from the
vector interaction in \eq{HV}, the net result as compared 
to the massive hydrogenic case is that the electronic mass is replaced
by the reduced mass in positronium (a difference by a factor of 2).
Since the E2 channel is totally suppressed, the next contribution comes
from the E3 channel giving rise to a correction to order $\a^4$ ($\a$
being the fine structure constant) to the leading 2E1 decay amplitude.
In contrast, on expanding the factor 
$ e^{ i \bk \cdot \br /2} - e^{i \bk \cdot \br /2}$ in the effective 
scalar photon interaction in \eq{HS}, we see that the leading nonzero
interaction becomes
\be
H_{S} \sim H_{S}^{(1)} \equiv - \sum_{(\bep, \bk)}
\sqrt{\frac{2\pi}{\om}} \cdot \frac{e \bP \cdot \bep}{2 m_e} 
\, e^{i \bk \cdot \bR} (i \bk \cdot \br) a_{(\bep, \bk)} +h.c. \ .
\label{eff}
\ee
We notice that as far as the atomic coordinate is concerned, the
interaction in \eq{eff} behaves like an E1 (electric dipole)
interaction whose coupling strength depends on the center of mass
velocity. For comparsion, we also approximate the vector interaction
$H_V$ of \eq{HV} by its leading E1 interaction:
\be
H_{V} \sim H_{V}^{(1)} \equiv  - \sum_{(\bep, \bk)}
\sqrt{\frac{2\pi}{\om}} \cdot \frac{e \bp \cdot \bep}{m_e /2} 
\, e^{i \bk \cdot \bR} a_{(\bep, \bk)} +h.c.
\ee
Both interactions $H_S^{(1)}$ and $H_V^{(1)}$ couple to the $|nP>$
state, but $H_S^{(1)}$ has a nonzero coupling to the $|2P>$ state 
whereas $H_V^{(1)}$ does not because of the $|2S>$ and $|2P>$
degeneracy. Thus the $|2P>$ state contributes to the two scalar photon
decay of metastable postitronium, but not to the two vector photon decay
nor to the one vector--one scalar photon decay. This follows from the
trivial identity
\be
<nP| \bp |2S> = \frac{i m_e }{2} E_{n2} <nP| \br |nS> , 
\ee
where $E_{n2}$ is the energy difference between the $|nP>$ and 
$|2S>$   positronium states and 
we have explicitly used the fact that the reduced mass in
positronium is $m_e/2$. Thus we see that the contribution of
$H_S^{(1)}$ to the two photon decay amplitude relative to that of 
$H_V^{(1)}$ is of the order
\be
\frac{<H_S^{(1)}>}{<H_V^{(1)}>} \sim \frac{1}{4} \frac{k P}{m_e E_{n2}}
\sim \frac{P}{m_e} \sim \frac{v}{c} ,
\ee
where $v$ is the velocity of the positronium center of mass motion, and
we have explicitly reinserted $c$. This simple analysis leads us to
conclude that
\be
\frac{M_{SS}}{M_{SV}} \sim \frac{M_{VS}}{M_{VV}}\sim \frac{v}{c} .
\ee

In this brief report, we consider a general hydrogenic system and study 
the effect of the scalar photon
interaction $H_S$ on the two--photon transition between two 
$S$-states ($nS$ to $n'S$). Thus, the general matrix element is of the
form 
\bea
&&<n'S| H_{sg} |nS>   \quad\quad \mbox{or} \nn\\
&&<n'S|H_i G H_j |nS>,\quad\quad 
\mbox{ where $H_{i(j)}$ can be either $H_V$ or $H_S$,}
\eea
and $G$ is the hydrogenic Coluomb Green's function. 
For example, $<n'S|H_V G H_V |nS>$ is the usual two vector photon
transition matrix element between $|nS>$ and $|n'S>$. This
can be calculated in closed form, following a method used by Gavrila and
Costescu \cite{ga}. More interesting are matrix elements of the type 
\bea
&&<n'S|H_S G H_V |nS>, \quad\quad 
\mbox{the mixed vecor--scalar matrix element,}\nn\\ 
&&<n'S|H_S G H_S |nS>, \quad\quad 
\mbox{the two scalar photon transition matrix element}.
\eea

The two scalar photon transtion matrix element for the special case
where $|nS>$ and $|n'S>$ are both the $ |1S>$ ground state has in fact
been evaluated analytically by one of us \cite{au2}. It is interesting
to point out that in the calculation reported in \cite{au2}, the two
scalar photon transition matrix element was only used as a calculational
tool and no physical significance was attached to it. The generalization
to arbitrary $n$ and $n'$ can be easily carried out and the result
will be reported in section 2 below. The mixed scalar--vector two photon
transition matrix element has never 
been studied before. It can also be evaluated in closed form and will 
be reported in section 2  below. In section 3, we apply these results to study the two--photon decay of
the metastable $2S$ state. The angular distribution of the two--photon
decay spectrum is reported. 

\section{Exact two photon matrix elements}

We will see in the following that for a scattering process
\footnote{Notice that for a two photon decay process, one only need to
replace $\bk$ by -$\bk$ and $\om$ by -$\om$ in \eqs{mvv}{mvs}. } from 
$(\bep, \bk)$ to $(\bep', \bk')$, it is useful to introduce the
the following two photon matrix elements:

the two vector photon matrix element,
\bea
M_{VV}(\Om,  \bk_2, \bk_1)
\equiv 
\int\int d\br_1 d\br_2 \;
\bep' \cdot \bp u_{n'S} (\br_2)  \: 
e^{-i  \bk_2 \cdot \br_2} G(\br_2, \br_1, \Om) 
e^{i \bk_1 \cdot \br_1} \:
\bep \cdot \bp u_{nS} (\br_1) \nn\\
= 
\int\int d\bp_1 d\bp_2 \;
\bep' \cdot \bp_2  u_{n'S} (\bp_2 -\bk_2)  \:
G(\bp_2, \bp_1, \Om) \:
\bep \cdot \bp_1  u_{nS} (\bp_1 -  \bk_1) ,
\label{mvv}
\eea  
the two scalar photon transition matrix element 
\be
M_{SS}(\Om,  \bk_2,  \bk_1) \equiv
\int\int d\bp_1 d\bp_2 \;
u_{n'S} (\bp_2 - \bk_2) \:
G(\bp_2, \bp_1, \Om) \:
u_{nS} (\bp_1 -  \bk_1),
\label{mss}
\ee
and the mixed scalar--vector two photon transition matrix element 
becomes
\bea
&&M_{SV}(\Om,  \bk_2,  \bk_1) \equiv
\int\int d\bp_1 d\bp_2 \;
u_{n'S} (\bp_2 - \bk_2) \:
G(\bp_2, \bp_1, \Om) \:
\bep \cdot \bp_1  u_{nS} (\bp_1 -  \bk_1), 
\label{msv}
\\
&&M_{VS}(\Om,  \bk_2, \bk_1) \equiv
\int\int d\bp_1 d\bp_2 \;
\bep' \cdot \bp_2  u_{n'S} (\bp_2 - \bk_2) \:
G(\bp_2, \bp_1, \Om) \:
u_{nS} (\bp_1 -\bk_1) . 
\label{mvs}
\eea
Here $ G(\bp_2, \bp_1, \Om)$ is the Coulomb Green's function 
$(\Om - H_0)^{-1}$
in momentum representation, 
$\bk_1 = -\d \bk$ or $(1-\d) \bk$ and 
$\bk_2 = \d \bk'$ or $- (1-\d) \bk'$ 
originating from the finite mass effect. 

To simplify the discussion, we limt ourself to a hydrogenic atom where
an electron is bound to a positive charge $+Z|e|$. We take advantage of
the fact that the $|nS>$ wave function can be generated from the $|1S>$
wave function by a parametric differentiation:
\be
\psi_{nS}(p) = -\frac{1}{n^{7/2} n!} (\frac{8 \l^5}{\pi^2})^{1/2}
B_n(-\frac{2\l}{n} \del_\b) \frac{1}{(p^2+\b^2)^2} |_{\b = \l/n} .
\label{wave}
\ee
where 
\be
B_n(x) \equiv L_n^1(x) -2 L_n^2(x), \quad\quad B_1(x) =1,
\ee
and $L_\mu^\nu(x)$ are the associated Laguerre functions. In  
\eq{wave}, $\b, n, \l$ are independent variables and 
\be
\l \equiv \a Z \mu.
\ee 
Here $\mu$ is the reduced mass of the electron in the atom. We also 
use natural units where $c = \hbar =1$.

Next we used the fact \cite{ga} that the integral 
\be
I(\Om, \bk_2, \bk_1) \equiv 
\int\int d\bp_1 d\bp_2 \;
((\bp_2 -\bk_2)^2 + \eta^2)^{-1} \:
G(\bp_2, \bp_1, \Om) \:
((\bp_1 -\bk_1)^2 + \b^2)^{-1} 
\label{I}
\ee
can be evaluated analytically in closed form. The analytical form of $I$
is given in  \eq{J2} below. Then a comparison of \eq{wave}
and \eq{I}, upon noticing the property of tranversality of the photon,
clearly suggests that to generate the needed matrix elements, we need
the simple relations:
\bea
&&-\frac{1}{2 \b} \frac{\del}{\del \b} \frac{1}{((\bp +\bk)^2 + \b^2)} 
= \frac{1}{((\bp +\bk)^2 + \b^2)^2}, 
\label{rel1}\\
&&-\frac{1}{4} \e_i \frac{\del}{\del k_i} \frac{1}{((\bp +\bk)^2 + \b^2)} 
= \frac{\bep \cdot \bp}{((\bp +\bk)^2 + \b^2)^2}.
\label{rel2}
\eea
Thus, the relations expressed in eqs. \eq{rel1} and \eq{rel2} become the
keys to generate a scalar photon vertex and a vector photon vertex
respectively.

Having laid down the rules to obtain the transition matrix elements, we
proceed to simply stating the results. We define the following:
\bea
&&X \equiv i (2\mu)^{1/2} (E_{ns} + \Om)^{1/2}, \\
&&E_{ns} \equiv -\l^2/(2\mu) n^2,\\
&&\tau \equiv \l/X, \\
&& \b \equiv \l/n , \quad\quad \eta \equiv \l/n',\\
&&s \equiv \frac{2 (\c^2 \bk^2 +\b^2 -X^2) ( \bk_2^2 +\eta^2 -X^2) 
- 8 X^2  \bk_1 \cdot \bk_2}
{[(X+\b)^2 +\bk_1^2][(X+\eta)^2 +\bk_2^2]} ,\\
&&p \equiv \frac{ ((X-\b)^2 +\bk_1^2) ((X-\eta)^2 +\bk_2^2)} 
{[(X+\b)^2 +\bk_1^2][(X+\eta)^2 +\bk_2^2]} 
\eea
and
\bea
J(\Om, \bk_2, \bk_1) &\equiv& 
64 \l \mu X (\frac{i e^{i \pi \tau} }{ 2 sin \pi \tau})
\int_1^{0_+} d \rho \rho^{-\tau} \frac{(1-s\rho +p \rho^2)^{-1}}
{((X+\b)^2 +\bk_1^2) ((X+\eta)^2 +\bk_2^2)} 
\nn\\
&=& \frac{8 \l^5}{\pi^2} I(\Om, \bk_2, \bk_1) , 
\label{J2}
\eea
where the integration contour begins at $\rho=1$, runs along the real
axis to a point closely on the right of $\rho=0$, encirlces the origin 
in the counter--clockwise sense and runs back to $\rho=1$.  
The integral $J$ is expressible in terms of the Apelle type
hypergeometric function. The desired matrix elements \eqs{mvv}{mvs}
are then expressible in terms of parametric differentiation
of $J$. Specifically,
\bea
&&M_{VV} = \frac{1}{16} \e'_i \e_j \frac{\del}{\del k_{2i}}
\frac{\del}{\del k_{1j}} J(\Om, \bk_2, \bk_1) ,\\
&&M_{SS} = \frac{1}{4} \frac{1}{\b \eta} \frac{\del}{\del \b}
\frac{\del}{\del \eta} J(\Om, \bk_2, \bk_1) ,\\
&&M_{SV} = \frac{1}{8} \frac{1}{\eta}\frac{\del}{\del \eta}
\e_i \frac{\del}{\del k_{1i}} J(\Om, \bk_2, \bk_1) ,\\
&&M_{VS} = \frac{1}{8} \e'_i \frac{\del}{\del k_{2i}}
\frac{1}{\b}\frac{\del}{\del \b } J(\Om, \bk_2, \bk_1).
\eea
This completes the analytical calculations of the two--photon transition
matrix elements. We would like to point out that the parametric
differentiation using 
$\e_i \del /\del k_i$ or $\e'_i \del /\del k'_i$
is rather simple upon using the tranversality condition
\be
\bep \cdot \bk = \bep' \cdot \bk' =0.
\ee

\section{Two photon decay of the metastable $2S$ state}

In this section, we consider the two photon decay of the metastable $2S$
state of a hydrogenic system with a electron bound to a positive charge
$Z |e|$. 
We will present all the results  consistently up to
order $\De /M$ and $(Z\a)^2$, compared to the leading order. 

\subsection{Kinematics}
Let the emitted photons be in the state 
$(\bep, \bk)$ and $(\bep',\bk')$. 
Without loss of generality, we can take the atom to be
initially at rest ($\bP=0$). From the conservation of 
energy and momentum, we have
\bea
&&\om +\om' + \frac{\bP'^2}{2 M} = \De, \nn\\
&&\bP' = -\bk - \bk' \ ,
\eea
where  $\De = \frac{3}{8} \mu Z^2 \a^2$ is the energy difference between
the $2S$ and $1S$ level.
This gives
\be
\om +\om' + \frac{\om^2 +\om'^2 + 2 \om \om' cos \th}{2M} = \De ,
\qquad \mbox{where $cos \th = \hat{\bk} \cdot \hat{\bk'}$}
\label{oo}
\ee
A convenient way to parametrize this equation is to introduce 
a dimensioness parameter $\s$ such that
\bea
&&\om =\De [ \s  +\frac{\De \s (1-\s)}{2M} (1-cos \th) 
-\frac{\De}{2M} \s +o(\a^3)] , \nn\\
&&\om' =\De [ (1-\s)  +\frac{\De \s (1-\s)}{2M} (1-cos \th) 
-\frac{\De}{2M} (1-\s) +o(\a^3)] .
\label{para}
\eea
This parametrization satisfies \eq{oo} up to order $\De/M$. 
The range of $\om$ is from 0 to $\De - \frac{\De^2}{2M}$ and
the range of $\s$ is from 0 to 1.

\subsection{Transition amplitudes}
The transition operator is given by
\be
T = H_1 + H_{sg} + H_1 \frac{1}{E-H_0} H_1 + \cdots ,
\qquad \mbox{where $H_1 \equiv H_V+H_S$}.
\label{T}
\ee
It is clear that the matrix 
element for the terms in ellipsis in \eq{T}
are of order higher than $(Z\a)^2$ or $\De/M$ and we only need to compute 
$<f| T |i>$ for the displayed terms in \eq{T}.

The initial and final states are 
\bea
|i> = |\bP> \otimes &|2S>  \otimes &
|\mbox{no photon} > , \nn\\
|f> = |\bP'> \otimes &|1S> \otimes &
|\mbox{two photons in the modes $(\bep, \bk), (\bep', \bk')$}> .
\eea
It is easy to see that 
\be
<f| H_1 |i> =0,
\ee
\be
<f| H_{sg} |i> = \frac{e^2}{m_2}
\{ R(1- \d) + Z^2 \frac{\d}{1-\d} R(\d) \} (\bep \cdot \bep') 
e^{-i (\bk +\bk')\cdot \bR},
\ee
where
\be
R(\d) \equiv  <1S| e^{-i \d (\bk+\bk') \cdot \br} |2S> ,
\ee
which to order $(Z\a)^2$ can be written as
\be
R(\d)= R_1(\d) + R_2(\d) cos \th,
\ee
where
\bea
&&R_1(\d) = \frac{256\sqrt{2}}{729}
\{ (\frac{3\s}{8})^2 + (\frac{3 (1-\s)}{8})^2 \} (Z \a)^2 \d^2,  \nn\\
&&R_2(\d) = \frac{8 \sqrt{2}}{81}
\s (1-\s) (Z \a)^2 \d^2 ; 
\eea
and
\be
<f|  H_1 \frac{1}{E-H_0} H_1 |i> =
e^{-i (\bk +\bk')\cdot \bR} 
( A(\bk, \bk') +B(\bk, \bk')+ (\bk \leftrightarrow \bk') ),
\ee
where $A$ comes from the two vector photon interaction,
$B$ comes from the mixed vector--scalar photon interaction. They are:
\bea
&&A(\bk, \bk') = -e^2 <1S| \bp \cdot \bep' 
( \frac{Z}{m_2} e^{i (\d -1) \bk' \cdot \br} +
  \frac{1}{m_1} e^{i \d \bk' \cdot \br} )
\Gh(\Om) 
(\frac{Z}{m_2} e^{i (\d -1) \bk \cdot \br} +
  \frac{1}{m_1} e^{i \d \bk \cdot \br} )
\bp \cdot \bep |2S> \nn\\
&&B(\bk, \bk') = -e^2  \frac{\bP' \cdot \bep}{M} <1S|
( Z_1 e^{i \d \bk' \cdot \br} +   Z_2 e^{i (\d-1) \bk' \cdot \br} ) 
\Gh(\Om) 
(\frac{Z}{m_2} e^{i (\d -1) \bk \cdot \br} +
  \frac{1}{m_1} e^{i \d \bk \cdot \br} )
\bp \cdot \bep |2S> \nn
\eea
It is clear that these matrix elements can be reduced to the forms
\eq{mvv} and \eq{mvs}.

In particular, we have (for $n=2, n'=1$),
\bea
M_{VV}  =&&  \frac{32 \sqrt{2} \mu \l^5 X^3}{f_1^3 f_2^2}
[ \; (X^2-\b^2 +\bk_1^2) \frac{F_A(2-\t;2,2;3-\t;x_1,x_2)}{2-\t} 
\label{r1} \\
  && - \b y_1 \frac{F_A(3-\t;3,3;4-\t;x_1,x_2)}{3-\t} 
   -\b y_1 \frac{F_A(4-\t;3,3;5-\t;x_1,x_2)}{4-\t} \; ]
     (\bep\cdot\bep')\nn\\
&&+\frac{512\sqrt{2}\mu \l^5 X^5}{f_1^4 f_2^3}
[ \; (X^2 -2\b^2 -X \b +\bk_1^2) 
\frac{F_A(3-\t;3,3;4-\t;x_1,x_2)}{3-\t}\nn\\
 && -\frac{3}{2} \b y_1 \frac{F_A(4-\t;4,4;5-\t;x_1,x_2)}{4-\t} 
   - \frac{3}{2} \b y_2 \frac{F_A(5-\t;4,4;6-\t;x_1,x_2)}{5-\t}
\; ] (\bep\cdot\bk_2) (\bep'\cdot\bk_1) \nn
\eea
and
\bea
M_{VS}  =&& 
\frac{64 \sqrt{2} \mu \l^4 X^3}{f_1^2 f_2^3} (\bep\cdot\bk_2)
[ \; 2(X+\l)(1- \frac{2 \b(X+\b)}{f_1}) 
\frac{F_A(2-\t;2,2;3-\t;x_1,x_2)}{2-\t}\nn\\
&&+ (y_1'+\frac{4 Q_1 \b}{f_1}) 
\frac{F_A(3-\t;3,3;4-\t;x_1,x_2)}{3-\t} 
+(y_2'+\frac{4 Q_2 \b}{f_1})
\frac{F_A(4-\t;3,3;5-\t;x_1,x_2)}{4-\t}\nn\\
&&-\frac{3}{2} \frac{\b W_1}{f_1} 
\frac{F_A(5-\t;4,4;6-\t;x_1,x_2)}{5-\t} 
-\frac{3}{2} \frac{\b W_2}{f_1} \frac{F_A(6-\t;4,4;7-\t;x_1,x_2)}{6-\t}
\; ]
\label{r2}
\eea
where $F_A$ is the Apelle's type hypergeometric function, 
\bea
&&\b = \l /2, \qquad \l =\a Z \mu,  \\
&&X^2 = 2 \mu (|E_{2s}| + \om +\frac{\om^2}{2M} ),  \\
&& \t =\l/X, \\
&&f_1 =(X+\b)^2 +\bk_1^2, \qquad  f_2 =(X+\l)^2 +\bk_2^2, \\
&&\ft_1=(X-\b)^2 +\bk_1^2, \qquad  \ft_2 =(X-\l)^2 +\bk_2^2, \\
&&g_1 =\bk_1^2 +\b^2 -X^2, \qquad g_2 =\bk_2^2 +\l^2 -X^2,  \\
&&y_1 = \frac{1}{f_1 f_2}
[	-4 g_2 X [(X+\b)^2 -\bk_1^2] 
	+16 X^2 (X+\b)(\bk_1\cdot\bk_2)
],\\
&&y_2 =\frac{1}{f_1 f_2}
[	4 X ( (X-\l)^2 + \bk_2^2 ) ( \b^2 -X^2 -\bk_1^2 ) 
],\\
&&Q_1 = \frac{1}{f_1 f_2}
[	-\l \b f_1 f_2 +3 \b(X+\l) f_1 g_2 
	+ 3 \l (X+\b) f_2 g_1
	-5(X+\b)(X+\l) g_1 g_2 	\nn\\
&&\qquad \qquad 
 	-20 X^2(X+\b)(X+\l) \bk_1\cdot \bk_2 
],\\
&&Q_2 =\frac{1}{f_1 f_2}
[	-4(X+\b)\l g_2 \ft_1 
      	-4(X+\l)\b g_1 \ft_2
      	+2 \l \b g_1 g_2
      	+6 \ft_1 \ft_2  (X+\b)(X+\l)
] \nn\\  	
&&\qquad \qquad 
	 -\frac{3}{8} y_1' y_2', \\
&&s = \frac{1}{f_1 f_2}
[	2 g_1 g_2 
     	+ 8 X^2  \bk_1 \cdot \bk_2 
], 
\qquad 
p = \frac{\ft_1 \ft_2}{f_1 f_2},  \\
&&W_1 =y_1 y_2'+ y_2 y_1' +s y_1 y_1', \qquad \qquad
W_2= y_2 y_2' -py_1 y_1'
\eea
and $x_1$, $x_2$ are given by
\be
1-s\rho + p\rho^2 =(1-x_1 \rho)(1-x_2\rho) .
\ee
Here $y_1', y_2'$ are obtained from $y_1, y_2$ by interchanging
$\l$ with  $\b$ and $\bk_1$ with $\bk_2$.

The $2S$ to $1S$ amplitude 
$M_{VV}$ was first evaluated in \cite{mvv1} within the dipole
approximation. Higher multipole contributions were calculated in
\cite{mvv2}. 
\footnote{In \eq{r1} above, we have corrected some typos in \cite{mvv2}.
In \cite{mvv2}, the total power of $f_1$ in the first (second) factor of
eqn. (7) should be 3 (4) instead of 2 (2);  in eqn.(9), it should read
$x_1* x_2$ instead of $x_1-x_2$; the L.H.S. of eqn. (13) should read 
$y_2$; the L.H.S. of eqn. (14) should read -$y_1$.
In \cite{mvv2}, there was an error (a factor of 2) in one of the terms
presented. The correct results are given in \eq{spec} and \eq{d0}  below.
}

One can substitute these expression into \eq{T} and obtain the 
exact two photon matrix elements. However, to illustrate the finite mass
effect, it is enough to do an expansion and keep results up to first
order in $\De/M$.

It is convenient to write $<2S|T|1S>$ in the form
\be
<f|T|i> = \frac{\a}{m_2}\frac{2 \pi}{\sqrt{\om \om'}} \cM.
\ee
Using \eq{para} and perform a tedious but systematic and straight 
forward expansion in power of $Z\a$ and $\De/M$, one obtain 
\be
\cM =(\hat{\bep}\cdot \hat{\bep'})
\{g + (Z\a)^2[f+  (f' +f'' \frac{\mu}{M}) cos \th ] \} 
+(Z\a)^2(h' + \frac{\mu}{M} h'' ) (\hat{\bep}\cdot \hat{\bk'})
(\hat{\bep'}\cdot \hat{\bk}),
\ee
up to order $(Z\a)^2$ and first order in $\De/M$.
We have singled out the $\d$ dependence and
$g, f, f', f''$ and $h', h''$ are $\d$ independent. 
Here $\th$ is the correlation angle between the two photons in the
initial rest frame of the decaying system.

The physical origin of these terms are:
$g$ corresponds to the two electric dipole transition 
without retardation; 
$f$ corresponds to correction to $g$ to order $(Z\a)^2$ due to
retardation effect;
$f', h'$ come from the two electric quadrupole transition. 
$f'$ also contains a contribution coming from the seagull term;
$f'', h''$ include the recoil correction to the lowest order in $\De/M$.

\subsection{Angular distribution of the decay spectrum} 

On substituting $m_2 $ by $m_e$, 
the  decay spectrum is given by
\be dW =\frac{1}{2\pi} \frac{\a^2}{m_e^2} \sum |\cM|^2
\om \om' d\om d cos\th .
\ee
Using \eq{para}, one finds that
\be
\om \om' d\om = \De^3 d \s \s (1-\s) 
	[1 -\frac{\De}{2M} (1+2 \s) -\frac{\De}{M} (1-\s) cos \th ] .
\label{phase}
\ee	
Subsituting $\De = \frac{3}{8} \mu (Z\a)^2   = \frac{3}{8} 
(1-\d) m_e (Z\a)^2 $,
one obtain
\bea
\frac{d W}{d\s d cos \th} = 
&& \frac{1}{2\pi}(\frac{3}{8})^2  Z^6  m_e \a^8  (1-\d)^3 \cdot \\ 
&&  \sum |\cM|^2  \s (1-\s) \cdot
	[ 1- \frac{3}{16}(1+2 \s) \frac{\mu}{M} (Z\a)^2 
	-\frac{3}{8} (1-\s) cos \th \frac{\mu}{M}  (Z\a)^2
	]  \nn
\eea
where the sum is over the polarizations of the two emitted photons.
To second order in $ Z\a$ and first order in $\De/M$, we have
\bea
\sum |\cM|^2 = && 
	[g^2 +2 g f (Z\a)^2] (1+ cos^2\th)
	+ (Z\a)^2 [ 2g (f'-h') +2 g (f''-h'') \frac{\mu}{M} ] 
	cos \th\nn\\
	+&&(Z\a)^2 [2g (f'+h') +2 g (f''+h'') \frac{\mu}{M} ] 
	cos^3 \th .  
\eea
Hence, one can write 
\bea
\frac{d W}{d cos \th} = && Z^6 [ 
	( d_1 +d_2 (Z\a)^2 + d_3 \frac{\mu}{M}(Z\a)^2 ) \cdot
	(1+cos^2 \th) \nn\\
	&&+(Z\a)^2 (d_4 + \frac{\mu}{M} d_5) cos \th +
	(Z\a)^2   (d_6 + \frac{\mu}{M} d_7) cos ^3 \th
    \; ]
\label{spec}
\eea
with
\be
\begin{array}{ll} 
d_1 = c \int d\s \s(1-\s) g^2, &
d_2 = c \int d\s \s(1-\s) 2 g f, \\ 
d_3 = - c\int d\s 3/16 \s(1-\s)(1+2\s) g^2, & 
d_4 = c\int d\s \s(1-\s) 2 g (f'-h'), \\
d_5 = c\int d\s 
[\s(1-\s) 2 g (f'' -h'') - 3/8 \s (1-\s)^2 g^2 ],&
d_6 = c \int d\s \s(1-\s) 2 g (f'+h'), \\
d_7 = c \int d\s 
[\s(1-\s) 2 g (f''+h'')-  3/8 \s (1-\s)^2 g^2 ] 
\end{array}
\ee
and $c = \frac{1}{2\pi}(\frac{3}{8})^2  m_e \a^8 \cdot (1-\d)^3$.

Integrating over $\s$  from 0 to 1 by using the Simpson rule with 500 divisions
and using $\frac{1}{2\pi}(\frac{3}{8})^2  m_e \a^8= 52.3951 
\; \mbox{sec}^{-1}$,
we obtain (all numerical coefficients below are in units of $\mbox{sec}^{-1}$)
\bea
&&d_i =d_{i0} (1-\d) (1-\d +Z\d)^4, \qquad 
	\mbox{for $i =1,3,5,7,$} \nn\\
&&d_2 = (1-\d) (1-\d +Z\d)^2 
	[ ((1-\d)^3+ Z^2 \d^3) a + (1-\d +Z\d) ((1-\d)^3+ Z \d^3) a'
	],\nn \\
&&d_4 = d_{40} (1-\d) (1-\d +Z\d)^2   ((1-\d)^3+ Z^2 \d^3)
	\nn\\
&&d_6= (1-\d) (1-\d +Z\d)^2 
	[((1-\d)^3+ Z^2 \d^3) b +  ((1-\d)^2 -Z \d^2) b'
	], 
\label{dgen}
\eea
with $a=- 4.3990 \times 10^{-1}$, $a'= -3.9877 \times 10^{-2}$
$b= -2.8271 \times 10^{-1}$, $b'= 7.2080 \times 10^{-2}$. 
The dependence on $\d$ in the coefficients $d_1$ through $d_7$
comes from the recoil effects in the two--photons phase space \eq{phase}.
In the infinite nuclear mass limit  $M= \infty$ and 
$\d$=0, the coefficients $d_i$'s
become $d_{i0}$ given by 
\bea
d_{10} = 3.0860, && d_{20} = -4.7978 \times 10^{-1},\nn\\
d_{30} = -1.1573, &&d_{40} =  -2.8271 \times 10^{-1},\nn\\
d_{50} = 3.4852 \times 10^{-1}, && d_{60} = -2.1063 \times 10^{-1},\nn\\
d_{70} = -6.1464 \times 10^{-1} .  
\label{d0}
\eea

For positronium, $Z=1$ and $\d=1/2$, the corresponding coefficients are
given by
\bea
d_1 = 1.5430, && d_{2} = -5.9972 \times 10^{-2},\nn\\
d_3 = -5.7865 \times 10^{-1}, &&d_{4} = d_6= -3.5339 \times 10^{-2},\nn\\
d_{5} = 1.7426 \times 10^{-2}, &&d_{7} = -3.0732 \times 10^{-1} .  
\label{dpost}
\eea
Notice that  the electric quadrupole moment 
vanishes for the positronium and 
{\it all} the $(Z\a)^2$ corrections in \eq{spec}
arise from finite mass correction. 
It is not suprising that $d_4=d_6$ for the positronium if we remember that 
$h'$ comes solely from the electric quadrupole transition  and is zero in 
the present case. 

\vspace{0.75cm}
{\Large \bf Acknowledgements}

We would like to thank C.K. Chow for collaboration in the initial stage of
this work and comments on the manuscript.  CKA acknowledges the
hospitality of the Chinese University of Hong Kong through a C.N. Yang
fellowship while on sabbatical leave from the University of South
Carolina.

\ed

The values of $\s (1-\s) g^2$, $2 \s (1-\s) g f$, 
$3/16 \s (1-\s)(1+2 \s) g^2$, $3/8 \s (1-\s)^2 g^2$,
$ 2\s (1-\s) g(f'-h')$,  $2\s (1-\s) g(f''-h'')$,
$ 2\s (1-\s) g(f'+h')$,  $2\s (1-\s) g(f''+h'')$
can be computed numerically.